\documentclass[a4]{spie}  %>>> use for US letter paper
\usepackage[]{graphicx, epstopdf}
\usepackage{float}
\usepackage{subfig}
\usepackage{amsmath}
\usepackage{amssymb}
\usepackage{mathtools}
\usepackage{array}

\title{Spatial Linear Dark Field Control: Stabilizing Deep Contrast for Exoplanet Imaging Using Bright Speckles
} 

\author{Kelsey Miller\supit{a,b}, Olivier Guyon\supit{a,b,c,d}, Jared Males\supit{a}
\skiplinehalf
\scriptsize
\supit{\scriptsize a}Steward Observatory, University of Arizona, 933 N. Cherry Ave, Tucson, AZ 85721 \\
\supit{\scriptsize b}College of Optical Sciences, University of Arizona, 1630 E. University Blvd., Tucson, AZ 85721 \\
\supit{\scriptsize c}Subaru Telescope, National Observatory of Japan, National Institutes of Natural Sciences, 650 N. A'ohoku Place, Hilo, HI 96720 USA\\
\supit{\scriptsize d}Astrobiology Center, National Institutes of Natural Sciences, 2-21-1 Osawa, Mitaka, Tokyo, JAPAN
}
\normalsize
\begin{document} 
  \maketitle 

%%%%%%%%%%%%%%%%%%%%%%%%%%%%%%%%%%%%%%%%%%%%%%%%%%%%%%%%%%%%% 

%>>>> Include a list of keywords after the abstract 

\keywords{exoplanet direct imaging, high contrast imaging, wavefront control}

%%%%%%%%%%%%%%%%%%%%%%%%%%%%%%%%%%%%%%%%%%%%%%%%%%%%%%%%%%%%%
\begin{abstract}
Direct imaging of exoplanets requires establishing and maintaining a high contrast dark field (DF) within the science image to a high degree of precision (10$^{-10}$).  Current approaches aimed at establishing the DF, such as electric field conjugation (EFC), have been demonstrated in the lab and have proven capable of high contrast DF generation.  The same approaches have been considered for the maintenance of the DF as well.  However, these methods rely on phase diversity measurements which require field modulation; this interrupts the DF and consequently competes with the science acquisition.    In this paper, we introduce and demonstrate spatial linear dark field control (LDFC) as an alternative technique by which the high contrast DF can be maintained without modulation.  Once the DF has been established by conventional EFC, spatial LDFC locks the high contrast state of the DF by operating a closed-loop around the linear response of the bright field (BF) to wavefront variations that modify both the BF and the DF.  We describe here the fundamental operating principles of spatial LDFC and provide numerical simulations of its operation as a DF stabilization technique that is capable of wavefront correction within the DF without interrupting science acquisition.    
\end{abstract}

\section{Introduction} \label{sec:intro}
In the last two decades, the existence of 3,498 exoplanets has been confirmed \cite{ExoplanetCount}, and in the upcoming era of 30 meter class ground-based telescopes and new space-based observatories, there is the promise of discovery, even characterization, of many more exoplanets, including potentially Earth-like worlds.   With such powerful capabilities on the horizon, it has become imperative to push stellar suppression technology to higher precision in order to directly image these exoplanets.  Combined with coronagraphy, current speckle nulling techniques \cite{Bottom2016_specklenulling} are capable of creating the dark field (DF) in the science image where light from an exoplanet orbiting its star can be detected.  In principle, these methods are capable of generating a DF with 10$^{-10}$ contrast that would enable the imaging of Earth-like exoplanets,\cite{Trauger2007_Exoplanets} but to generate and maintain a DF at such contrast requires high precision measurements of very small changes in the science image. \\

The DF is very susceptible to small dynamic aberrations in the beam path.  Aberrations produce post-coronagraph stellar light leakage and result in a quasi-static speckle field in the image plane that limits the DF contrast. \cite{Traub2010_Exoplanets}.  Conventional wavefront sensors (WFS) that branch off from the main beam path cannot correct for non-common path (NCP) errors.  The resulting quasi-static aberrations can eventually become larger than the residual dynamic wavefront errors after correction by the WFS control loop.  For high contrast imaging, the WFS should ideally be common path with all of the optics seen by the science instrument to allow access to all aberrations created in the science beam.  This is achieved by focal plane wavefront sensing (FPWFS) which uses the science detector as the WFS to measure the exact aberrations seen by the main science beam.  \cite{Guyon2005_AOLimits}$^{,}$\cite{Delorme2016_FPWFS_SelfCoherentCam}  FPWFS techniques like speckle nulling and EFC that have proven capable of generating a DF with high contrast in the lab have also been under consideration for maintenance of the DF. \cite{Cady2013_EFCdemo}$^{,}$\cite{Ruffio2014_NCPCorrection}$^{,}$\cite{Krist2015_WFIRST_EFC}  As a control method, FPWFS presents its own set of challenges given that speckles have a quadratic relationship with aberrations.  These techniques also rely on phase diversity measurements of the field at the science detector which require field modulation and multiple images. \cite{Giveon2007_BroadbandHighContrast}$^{,}$\cite{Groff2015_FPWFS}  This field modulation at the science detector, induced by a deformable mirror (DM), throws stellar light back into the DF and disrupts the science measurement.  This interruption, which is required to rebuild the DF every time the contrast degrades, fundamentally limits the integration time that can be spent on any given target.  The duration of this interruption to the science acquisition is directly related to the contrast of the DF.  For deeper contrast, the required exposure time to sense the speckle field increases; therefore, at the 10$^{-10}$ contrast level, multiple images with long exposure times significantly reduces the amount of time that can be spent on observations.  The need for modulation, multiple images, and long exposures, consequently makes the use of current speckle nulling methods and EFC non-ideal for continuous maintenance of the DF.  \\
 
Another technique known as the self-coherent camera (SCC) has been under development as a method for obtaining and maintaining the DF without science acquisition competition.\cite{Delorme2016_FPWFS_SelfCoherentCam}  While SCC does not require modulation, it does still utilize the mixing of some starlight with the DF.  Linear dark field control (LDFC) does not require any such mixing of starlight and the DF, and offers a potential solution for overcoming the limitations presented by speckle nulling and EFC.  To avoid disrupting the science measurement with field modulation to rebuild the DF, LDFC locks the high contrast state of the field once the DF has been constructed using conventional methods like EFC.  Using only one image of the bright field (BF), LDFC freezes the state of the field by sensing and canceling changes in the wavefront that result in speckle formation in the image plane.  The ability to maintain the DF with a single image yields a substantial increase in time that can be spent in the observation and analysis of exoplanets and will lead to an overall increase in the number of planets detected and analyzed over the lifetime of an instrument.  \\

%------------------------------------------------
\section{Concept} \label{sec:concept}
LDFC maintains high contrast without needing to modulate the field and interrupt the science measurement to update the field estimate as is required when using EFC in closed loop.  Conceptually, LDFC is inspired by low-order wavefront sensing (LOWFS), a proven technique designed to actively sense and correct low-order wavefront aberrations using light rejected by a reflective stop placed in a pupil plane \cite{Singh2015_LOWFS} or focal plane \cite{Guyon2009_CLOWFS}$^{,}$\cite{Wallace2010_GPI} within the coronagraph.  LDFC is a common path FPWFS technique with the added advantage of access to mid- and high-spatial frequencies.  Instead of sensing only low-order aberrations using a post-coronagraph quadrant method or starlight rejected by the coronagraph, LDFC operates a closed-loop around starlight in the focal plane located outside of the DF.  \\

\begin{figure}[H]
\centering
\includegraphics[width=0.6\textwidth]{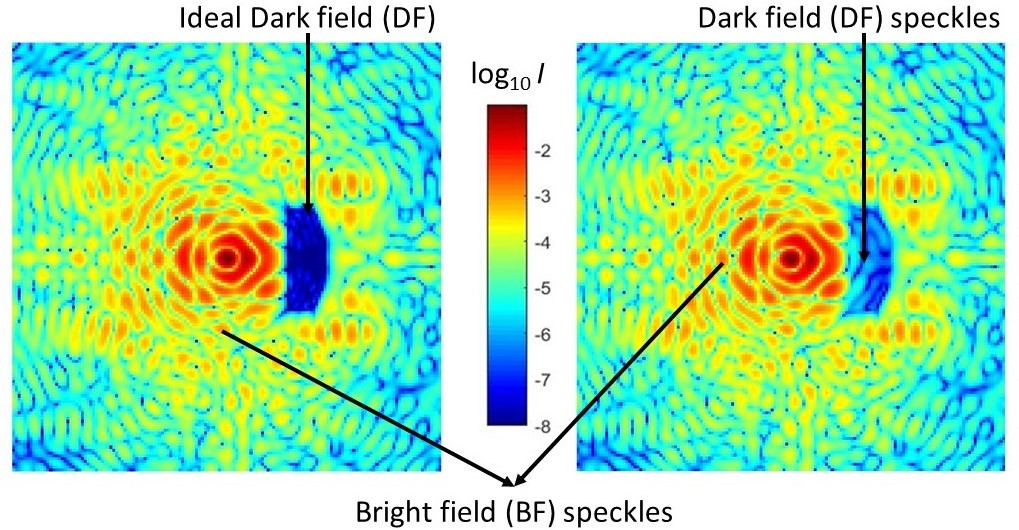}
\caption{Spatial LDFC: 3.5 $\lambda$/D x 8.5 $\lambda$/D DF created in simulation using conventional EFC (left).  Wavefront aberrations produce speckles in the BF and the DF which degrade the DF (right).  The change in intensity between the aberrated BF (right) and the ideal BF (left) is used to measure and cancel the speckles in the DF.  }
\label{fig:BFandDF}
\end{figure}

Spatial LDFC freezes the state of the DF by using state measurements of light spatially outside of the DF (see Fig \ref{fig:BFandDF}).  This method uses the linear signal from the strongly illuminated bright field (BF) to measure the change in the image plane intensity and uses that variation to calculate the correction required to return the image to its initial state, thereby stabilizing the DF.  Unlike other FPWFS techniques, spatial LDFC does not rely on any induced modulation to derive an estimate of the field to be canceled.  Instead, spatial LDFC observes changes in the image intensity with respect to a reference image taken after the DF has been established by conventional speckle nulling methods.  This process requires only the reference image and a single image taken at a later time.   \\

Without the need for modulation or multiple images, spatial LDFC does not interrupt the science measurement, it decreases the time necessary to return the DF to its initial high contrast state, and consequently it allows for longer, uninterrupted observing at high contrast.  This paper introduces spatial LDFC as a more efficient alternative to conventional speckle nulling methods for stabilizing the DF.   The theory behind spatial LDFC is laid out here in Section \ref{sec:concept}, and demonstrations of LDFC's abilities in simulation are shown in Section \ref{sec:results}.  Further discussion of the limitations and null space of spatial LDFC is laid-out in Section \ref{sec:discuss} with concluding remarks in Section \ref{sec:conclusion}. \\

\subsection{Theory} \label{sec:theory}
Spatial LDFC relies on the linear response of the BF to wavefront perturbations that affect both the BF and the DF; this linearity allows for a closed-loop control algorithm directly relating wavefront perturbations to changes in BF intensity.  To derive the source of this linear response, we begin with the relationship between an incident wavefront and the resulting image.  The complex amplitude of the incident wavefront in a pupil plane \textit{E$_{0}$} is linearly related to the complex amplitude at the image plane \textit{E$_{t}$} at a given time \textit{t}.  The same linear relationship is true with respect to \textit{E$_{DM}$}, the multiplicative complex amplitude introduced by the DM in a conjugate pupil plane, and the complex amplitude at the image plane \textit{E$_{t}$}. \\  

When the changes in optical path length (OPL) induced by the DM are very small  such that OPL $\ll$ 1, the resulting field \textit{E$_{t}$} at a given time \textit{t} in the image plane can be written as the sum of the initial pupil plane field \textit{E$_{0}$} and the small changes in complex amplitude induced in a conjugate pupil plane by the DM \cite{Giveon2007_BroadbandHighContrast}.

\begin{equation}
E_{t} \approx E_{0} + E_{DM}
\label{eq:Eimageplane}
\end{equation}

The resulting intensity in the image plane at time \textit{t} is then given by:     

\begin{equation}
I_{t} = |E_{t}|^{2}
\label{eq:intensity1}
\end{equation}

The total image plane intensity can be written as a sum of three terms: the intensity contribution from the initial pupil field: $\vert$\textit{E$_{0}$}$\vert$$^{2}$, the resulting intensity due to phase perturbations induced by the DM: $\vert$\textit{E$_{DM}$}$\vert$$^2$, and the inner product of the initial pupil field and the DM contribution to the complex amplitude:

\begin{equation}
	I_{t} \approx |E_{0}|^{2} +  |E_{DM}|^{2} + 2\langle E_{0} , E_{DM}\rangle
\label{eq:intensity2}
\end{equation}

In the DF, the contribution of the initial field to the total intensity is very small, and the total intensity is dominated by the contribution of the DM such that $|E_{DM}|^{2}$  $\gg$ $|E_{0}|^{2}$, thereby leading to a quadratic dependence of the DF on the DM input.  However, in the BF the contribution of the initial field to the total intensity dominates the contribution of the DM:

\begin{equation}
|E_{0}|^{2} \gg |E_{DM}|^{2}
\label{eq:assumption}
\end{equation}

The intensity of the BF at the image plane at time \textit{t} can therefore be written as a linear function of the complex amplitude contribution of the DM:

\begin{equation}
I_{t} \approx 2\langle E_{0} , E_{DM}\rangle + |E_{0}|^{2}
\label{eq:BFapprox}
\end{equation}

In Eq \ref{eq:BFapprox}, the term $|E_{0}|^{2}$ is the reference image \textit{I$_{ref}$} taken after the DF has been established.  The signal used by spatial LDFC to drive the DF back to its initial state is simply the difference between this reference and a single image \textit{I$_{t}$} taken at time \textit{t}.   

\begin{equation}
\Delta I_{t} = I_{t} - I_{ref} \approx 2\langle E_{0} , E_{DM}\rangle
\label{eq:lineardepend}
\end{equation}

\begin{figure}[H]
\centering
\includegraphics[width = 1\textwidth]{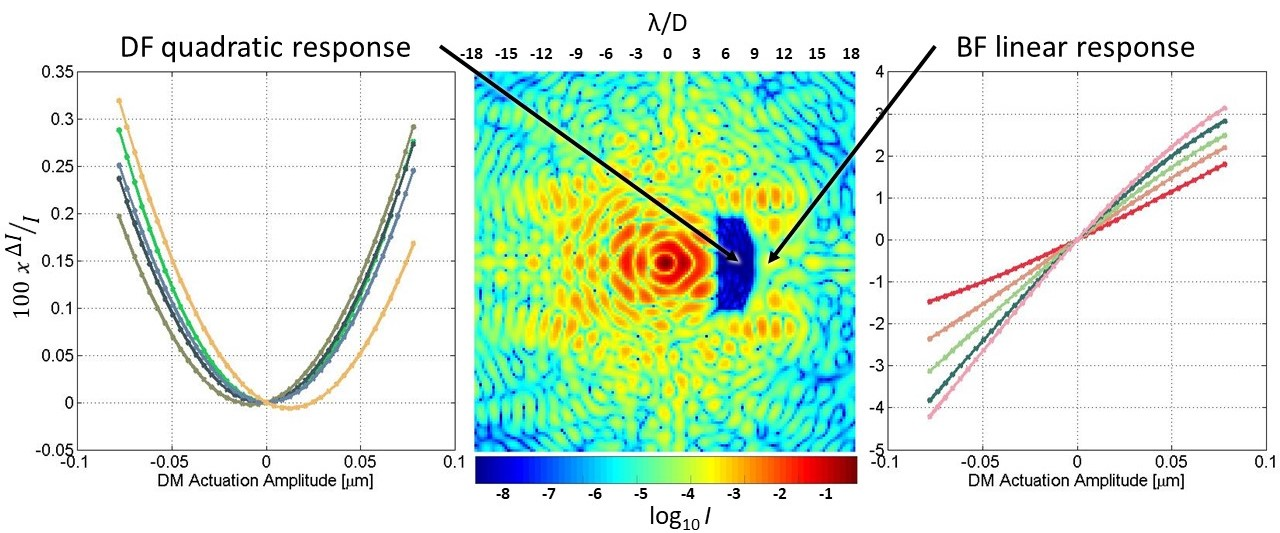}
\caption{The response of the BF and DF to the same range of DM 'poke' amplitudes from -0.075$\mu$m to +0.075$\mu$m on a single DM actuator.  A demonstration of the expected quadratic response of the DF (seen left) and the linear response of the BF (seen right) to the same range of DM poke amplitudes.  Each curve in the DF plot is the response of a single pixel in the DF between 7 $\lambda$/D and 8 $\lambda$/D, and each curve in the BF plot is the response of a single pixel in the BF between 10.5 $\lambda$/D and 11.5 $\lambda$/D.   }
\label{fig:pixelresponse}
\end{figure}

This linear response of the BF, \textit{$\Delta$I$_{t}$}, to field perturbations controlled by the DM is shown in Fig \ref{fig:pixelresponse} alongside the quadratic response of the DF to the same DM perturbation .  In this figure, the BF and DF response to the DM field contribution $E_{DM}$, is shown in a simulated PSF with a DF established by conventional EFC.  A model of a MEMS DM was used to create the DF and then perturb the input wavefront by inducing a positive and negative delay in the optical path with a single actuator.  This was done for a range of actuator amplitudes from -0.075 to +0.075 $\mu$m. \cite{Miller2016_LDFC}   The resulting intensity response of pixels located in the DF (shown to the left in Fig \ref{fig:pixelresponse}) is governed by Eq \ref{eq:intensity2} with the expected quadratic dependence on the field perturbation.  The intensity response of the BF (shown to the right in Fig \ref{fig:pixelresponse}) reveals the predicted linear dependence on the DM-induced field perturbation given by Eq \ref{eq:lineardepend} . \\      

In closed loop, \textit{$\Delta$I$_{t}$} is small, and the linear approximation holds.  However, even when initially closing the loop where \textit{$\Delta$I$_{t}$} is larger, strict linearity is not required, only a monotonic trend.  In instances both of strict linearity or of monotonicity, the BF response allows for the construction of a linear servo driven solely by changes in the BF intensity.  Unlike EFC and speckle nulling which use modulation to provide an absolute field measurement, LDFC relies on relative measurements of BF intensity variation in the science image.  These intensity variations are used to track and cancel changes in the wavefront that modify both the BF and DF, thereby stabilizing the DF without any disruptions to the science measurement.  \\  

\subsection{Calibration} \label{sec:calibration}
Given the linear relationship between BF intensity and wavefront, using LDFC to stabilize the DF contrast is faster and more robust than using EFC.  EFC requires multiple images to estimate the field, while each iteration of LDFC requires only one image at the science detector to determine how the field has changed with respect to the initial EFC-derived state.  Since this image does not require field probing which breaks the science measurement, the LDFC servo operates with a 100\% duty cycle.  Furthermore, LDFC does not rely on complex field estimates which require a model-based complex phase response matrix that is difficult to measure and verify; instead, LDFC relies only on a DM $\rightarrow$ image calibration that links a set of DM shapes, or basis functions, to changes in intensity in the science image. \cite{Guyon2015_WFSJPL}\\  

For this simulation, the DM influence functions were chosen as the basis functions.  The calibration between the image and the basis set was obtained by building a response matrix $\mathcal{M}$ whose columns relate the application of each individual influence function to the responding intensity variation at the science detector.  Though modal control \cite{Poyneer2005_modal_control} does offer performance benefits, especially when it maps with the expected temporal evolution of the wavefront error, such modal control tuning has not been explored at this time.  For this work, application of the influence function basis set involved the actuation or 'poking' of one of the \textit{k} actuators on the DM that lie within the illuminated system pupil.  To begin building $\mathcal{M}$, the ideal reference image \textit{I$_{ref}$} with dimensions [\textit{n$_{pix}$} x \textit{n$_{pix}$}] is recorded after the DF has been established using EFC.   To fill each of the \textit{k} columns in $\mathcal{M}$, a single actuator was poked, the resulting perturbed image \textit{I$_{k}$} was measured, and the unperturbed reference image \textit{I$_{ref}$} was subtracted off to yield the change in intensity.  This was done for all \textit{k} actuators.  The resulting response matrix  $\mathcal{M}$  has the dimensions [\textit{n$_{pix}^{2}$} x \textit{k}]. \\ 
\begin{equation}
\mathcal{M}[:\ , k] =  I_{k} - I_{ref}
\label{eq:calmat}
\end{equation}

\begin{figure}[H]
\centering
\qquad
\subfloat[BF pixel mask]{\includegraphics[width = 0.25\textwidth]{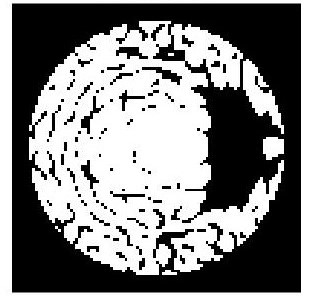}}
\qquad
\subfloat[\textit{I$_{ref,n}$}]{\includegraphics[width=0.26\textwidth]{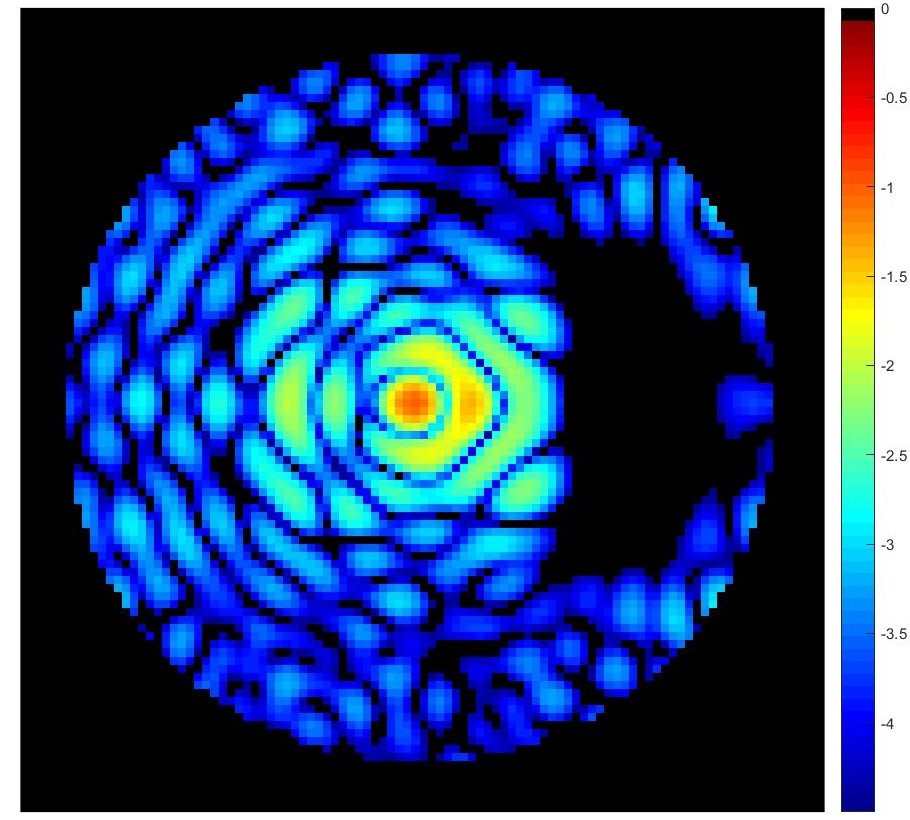}}
\caption{Images of the applied BF pixel mask (a) and the log$_{10}$ masked reference image (b).  The binary mask passes only the pixels at or above the contrast threshold (shown in white).  In this image, and for the following demonstrations, the contrast threshold was 10$^{-4.5}$, and the outer diameter of the masked control area was set to be the control radius of the active area on the DM.}
\label{fig:pixelsused}
\end{figure}

The matrix $\mathcal{M}$ records the intensity change of both the BF and DF pixels with respect to each actuator poke.  However, spatial LDFC uses only the BF pixels which respond linearly to wavefront perturbations.  The selection of these BF pixels relies on multiple parameters including background flux, flux per speckle, detector efficiency, and SNR.  Based on these requirements, a threshold was applied to the initial EFC image \textit{I$_{ref}$} which selected only the \textit{n} pixels with intensities greater than or equal to the threshold.  The result was an image \textit{I$_{ref,n}$} that recorded the initial EFC-state of only the BF pixels.  An example of this BF reference image and the corresponding pixel map can be seen in Fig \ref{fig:pixelsused} for a contrast threshold of 10$^{-4.5}$.  \\

To build the BF response matrix \textit{M}, the full response matrix was filtered to include only the \textit{n} BF pixels with intensities above the threshold such that \textit{M} = $\mathcal{M}_{n}$.  This filtered response matrix \textit{M} was used throughout the operation of spatial LDFC. \\

\subsection{Closed-loop implementation} \label{sec:implement}
\begin{figure}[H]
\centering
\includegraphics[width = 0.75\textwidth]{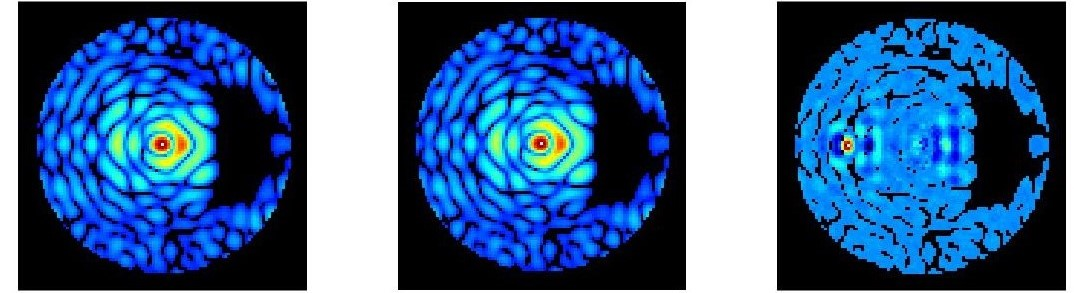}
\caption{Shown here are the BF pixels that are used in the reference (left) and aberrated (center) images to measure the intensity change (right) \textit{$\Delta I_{t,n}$} that drives the spatial LDFC control loop to stabilize the DF.  In all three images, the 3.5 $\lambda$/D x 8.5 $\lambda$/D DF can be seen to the right of the PSF core.}
\label{fig:deltaT}
\end{figure}

To implement LDFC in closed-loop, an image\textit{ I$_{t}$} was taken at time \textit{t} and the same \textit{n} BF pixels that pass the threshold were recorded in the BF image \textit{I$_{t,n}$}.  The BF reference image \textit{I$_{ref,n}$} was then subtracted from the new BF image to track the changes that occurred in the BF with respect to the initial EFC BF reference (see Fig \ref{fig:deltaT}): \\
\begin{equation}
\Delta I_{t,n} = I_{t,n} - I_{ref,n}
\label{eq:deltaI}
\end{equation}

This BF intensity change \textit{$\Delta$ I$_{t,n}$} was fit to the pseudo-inverse of the BF response matrix \textit{M}, also known as the control matrix, to calculate the DM shape that returned the field to its initial EFC reference state.  The DM shape is represented by a vector of individual actuator amplitudes \textit{u$_{t}$}:

\begin{equation}
u_{t} = -(M^{T}M)^{-1}M^{T}\ \Delta I_{t,n}
\label{eq:uk}
\end{equation}

This pseudo-inverse of \textit{M} was implemented by using singular value decomposition (SVD) and applying a threshold to filter out the modes that were not properly sensed by LDFC.  For this simulation, the threshold value was chosen based on simulation performance, resulting in the inclusion of 286 out of an initial 398 modes in the pseudo-inversion process.  A plot of the singular values of \textit{M} is shown below in Fig \ref{fig:SVDmodes}.    

\begin{figure}[H]
\centering
\includegraphics[width = 0.7\textwidth]{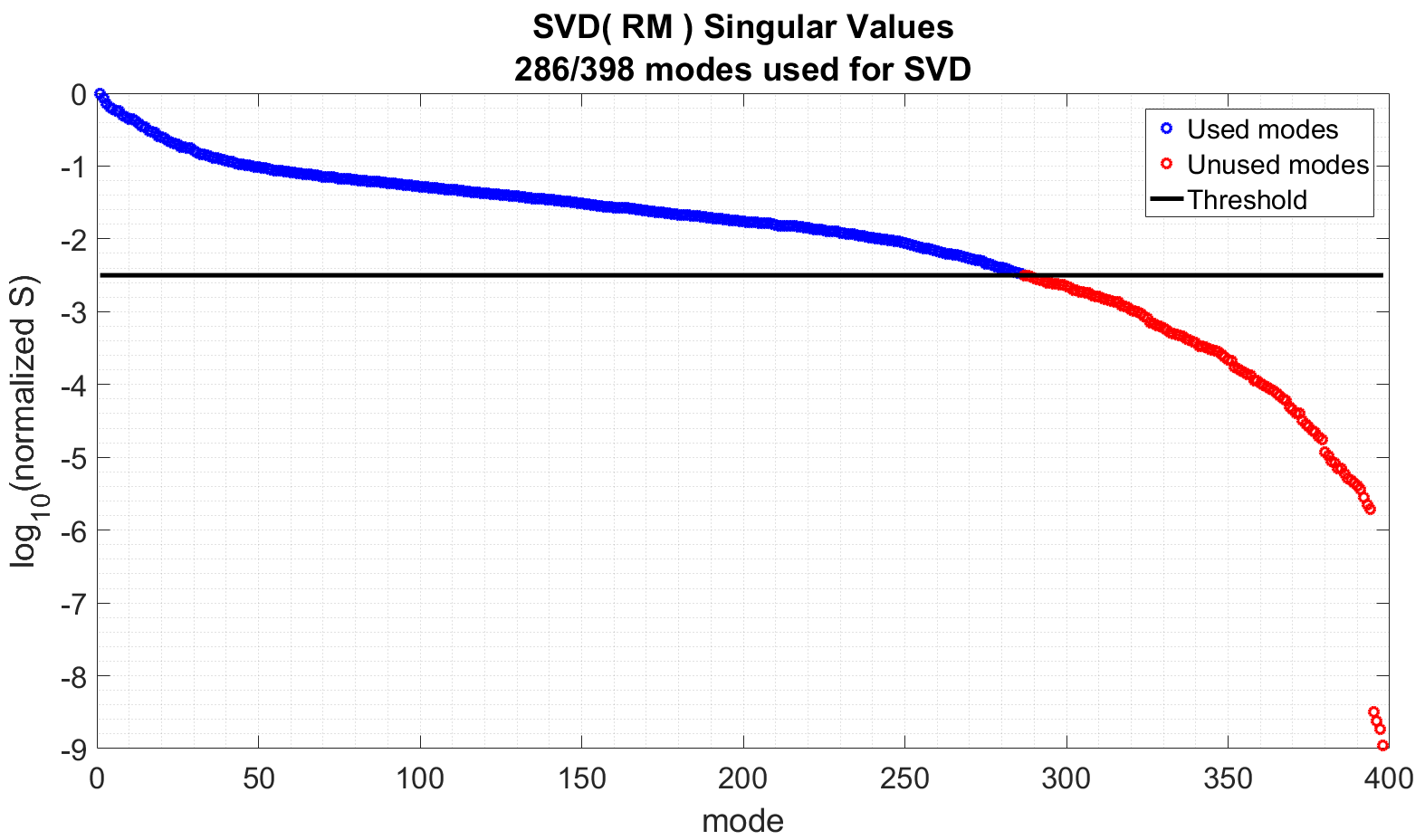}
\caption{Singular values of the spatial LDFC response matrix \textit{M} showing the applied SVD threshold (black) as well as the modes that were used in the inversion (blue) and the modes that were discarded (red).  Out of 398 total modes, 286 were used for the inversion of \textit{M} in the following simulations. }
\label{fig:SVDmodes}
\end{figure}

The response matrix \textit{M} and subsequent control matrix were measured once and applied in closed loop with an initial gain of 0.6.  Once the DF contrast converged to 10$^{-7.9}$, the gain was lowered to 0.1 to maintain the correction.  The ensuing process of taking an image, calculating the intensity change of the BF from its reference state, and updating the DM was iterated on to actively freeze the science image field in its initial EFC state. \\
%------------------------------------------------
\section{Simulation Results} \label{sec:results}
\begin{figure}[H]
\centering
\qquad
\subfloat[Telescope pupil and DM]{\includegraphics[width = 0.23\textwidth]{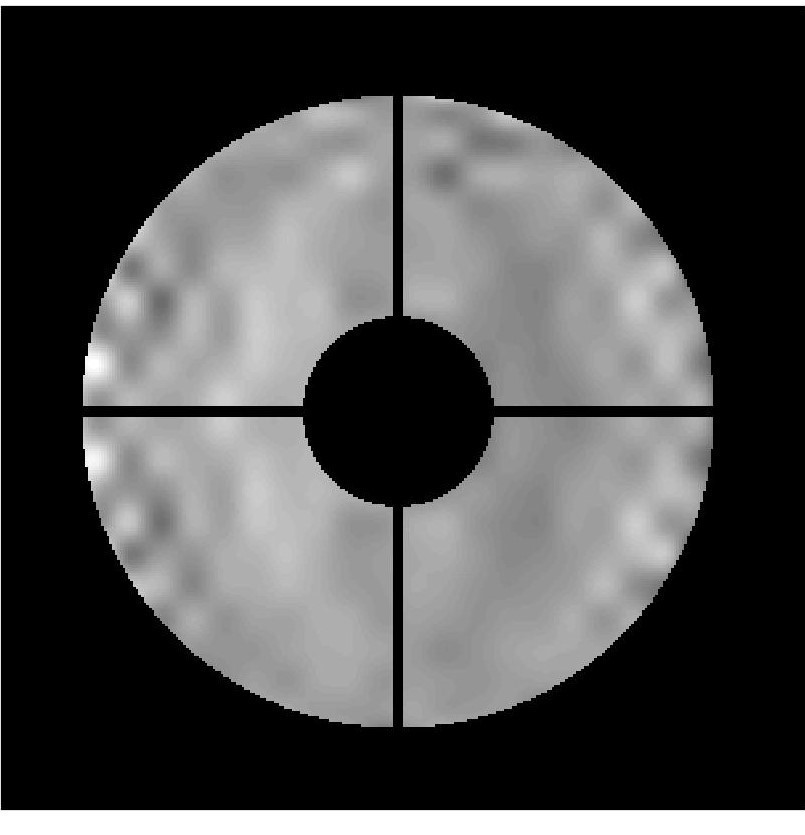}}
\qquad
\subfloat[Log$_{10}$ EFC reference image]{\includegraphics[width=0.28\textwidth]{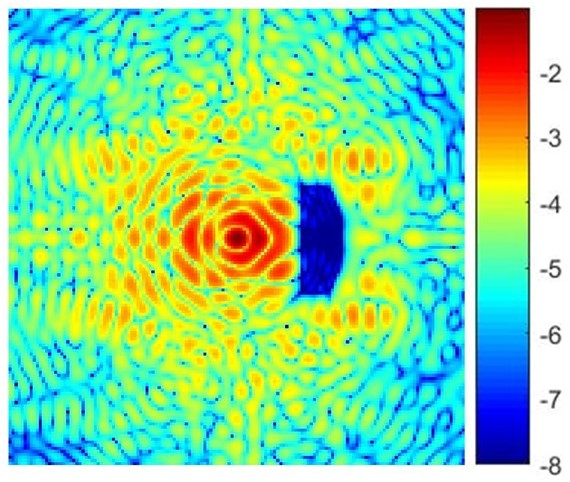}}
\caption{Standard implementation of EFC using a DM with 398 illuminated actuators (a) to create a 3.5 $\lambda$/D x 8.5 $\lambda$/D DF centered at 6.75 $\lambda$/D from the center of the stellar PSF with 10$^{-7.94}$ average contrast (b).}
\label{fig:EFCDH}
\end{figure}

To demonstrate spatial LDFC's ability to maintain the high contrast DF, a 6.5 m telescope system was constructed in simulation which includes a single DM and Lyot coronagraph that removes approximately two orders of magnitudes of stellar light from the final image.  The system entrance pupil was a 6.5 m diameter circular, centrally-obscured mask  with a 30$\%$ central obscuration and 2$\%$ spiders (see Fig \ref{fig:EFCDH}a).  The Lyot coronagraph consists of a Lyot stop undersized by 1$\%$ and a focal plane mask with a diameter of 2.44 $\lambda$/D.   For the system's DM, a model of a Boston Micromachines 1K DM was defined using 1024 actuators sharing a common gaussian influence function and 15$\%$ inter-actuator coupling.  The diameter of the illuminated pupil projected onto the DM was 6.5 mm, covering approximately 21 actuators and lending an outer working angle (OWA), or control radius of 10.5 $\lambda$/D.  Sampling at the science detector was 0.24 $\lambda$/D per pixel.  The source was a magnitude 5 star with sensing done at $\lambda$=550 nm (V band) with 10$\%$ bandwidth.  The total flux at the entrance pupil was 1.82x$10^{9}$ photons/second, and this rate was used to embed photon noise in all of the \textit{I$_{t}$} images in Eq \ref{eq:deltaI}.  All of these test parameters are listed in Table \ref{tabular:params}.
\begin{table}[H]
    \begin{center}
    \scriptsize
    \renewcommand{\arraystretch}{1.5}
    \begin{tabular}{| >{\centering\arraybackslash}m{1.5in} | >{\centering\arraybackslash}m{1.5in} |}
    \hline
    { Stellar magnitude} & { 5}   \\[0ex] \hline
    { Total flux} & { 1.82x$10^{9}$ photons/second}   \\[0ex] \hline
    { Noise included} & { photon noise}   \\[0ex] \hline
    { Exposure time} & { 5 seconds}   \\[0ex] \hline
    { Source wavelength} & { 550 nm, V band}   \\[0ex] \hline
    { Source bandwidth} & { 10$\%$}   \\[0ex] \hline
    { Telescope diameter} & { 6.5 m}   \\[0ex] \hline
    { Sampling at detector} & {0.24 $\lambda$/D per pixel}   \\[0ex] \hline
    { \# DM actuators used} & { 398, (21 in diameter)}   \\[0ex] \hline
    { \# Bright field pixels used} & {4535}   \\[0ex] \hline
    { Bright field contrast threshold} & {10$^{-4.5}$}   \\[0ex] \hline
    { Inner working angle (IWA)} & {2.44 $\lambda$/D}   \\[0ex] \hline
    { Outer working angle (OWA)} & {10.5 $\lambda$/D}   \\[0ex] \hline
  \end{tabular} 
  \caption{Simulated system parameters used in the following spatial LDFC demonstrations}
  \label{tabular:params}
  \end{center}
\end{table}
To build the DF, a standard implementation of EFC \cite{Groff2015_FPWFS} was used to suppress the stellar light to an average contrast floor of 10$^{-7.94}$ within a 3.5 $\lambda$/D x 8.5 $\lambda$/D region centered at 6.75 $\lambda$/D from the PSF core (shown in Fig \ref{fig:EFCDH}b).  This DF was the ideal reference state for LDFC to maintain, and the intensity image \textit{I$_{ref}$} was saved as the reference image to be used in the LDFC servo to return the DF to its EFC-derived state.\\
With the DF established, the spatial LDFC algorithm was implemented as described in Section \ref{sec:implement} to maintain the DF in the presence of two separate injected phase aberrations.  In the first case, a single speckle pair was induced in the image plane by applying a sine wave phase perturbation in the pupil.  For the second case, a random Kolmogorov phase screen was introduced in the pupil creating multiple speckles in the image plane.  In both cases, the same optical system, source, and threshold values were kept constant as were all other simulation parameters.    The following sections present spatial LDFC's response to these two cases.  \\
 
\subsection{Sine wave phase perturbation} \label{sec:linear}
\begin{figure}[H]
\centering
\includegraphics[width = 0.75\textwidth]{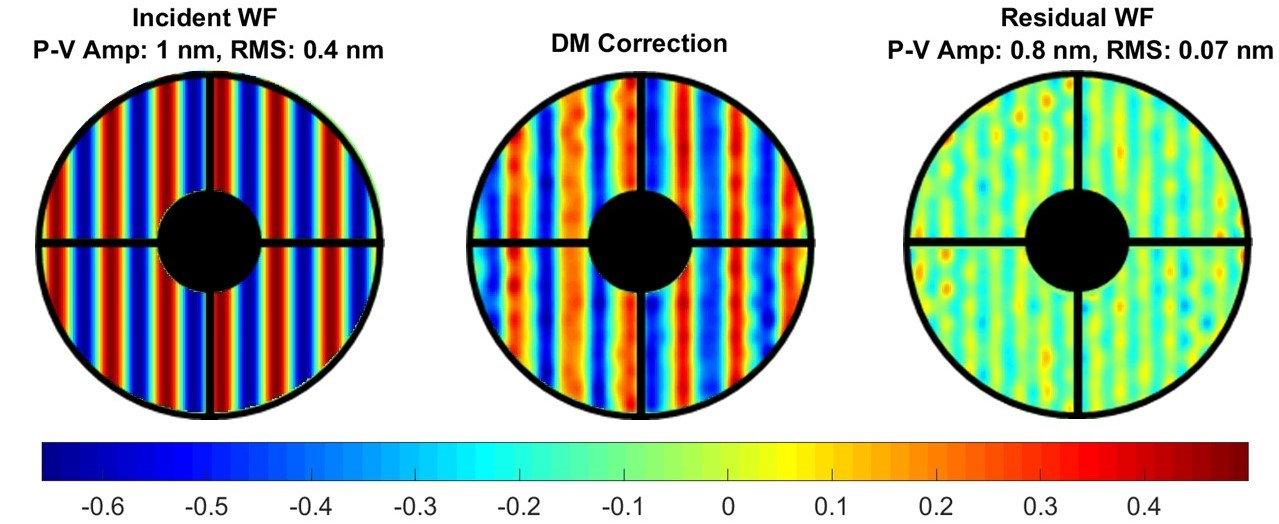}
\caption{The injected 6 cycles/aperture sine wave phase perturbation with a P-V amplitude of 1 nm (left), the DM response derived by LDFC (center) and the final residual wavefront error (right) after 6 iterations.  In this case, the residual WFE is dominated by a mode with a frequency of approximately 14 cycles/aperture which falls beyond the spatial frequency limit (10.5 cycles/aperture) the DM can correct.  This residual WFE is due to the gaussian shape of the DM's influence functions which cannot perfectly fit the injected sine wave perturbation, thereby leaving a residual sinusoidal pattern.  Scale is given in nm.}
\label{fig:speckle_full_display_PUPILs}
\end{figure}

\begin{figure}[H]
\centering
\includegraphics[width = 0.75\textwidth]{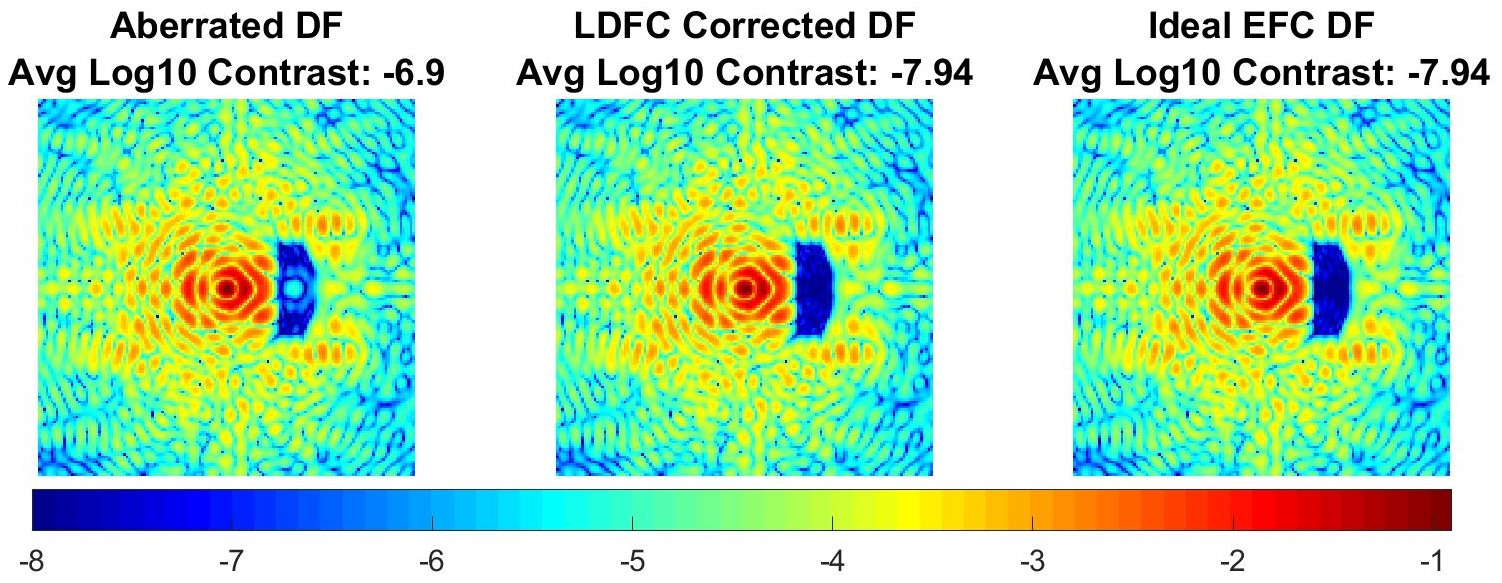}
\caption{The aberrated PSF with a single 10$^{-5.0}$ magnitude speckle in the DF with 10$^{-6.90}$ average contrast and a matching speckle in the BF, the final LDFC-corrected DF with 10$^{-7.94}$ average DF contrast, and the reference EFC-derived DF also with 10$^{-7.94}$ average DF contrast.  Scale is log$_{10}$ contrast.}
\label{fig:speckle_full_display_PSFs}
\end{figure}

\begin{figure}[H]
\centering
\includegraphics[width=0.75\textwidth]{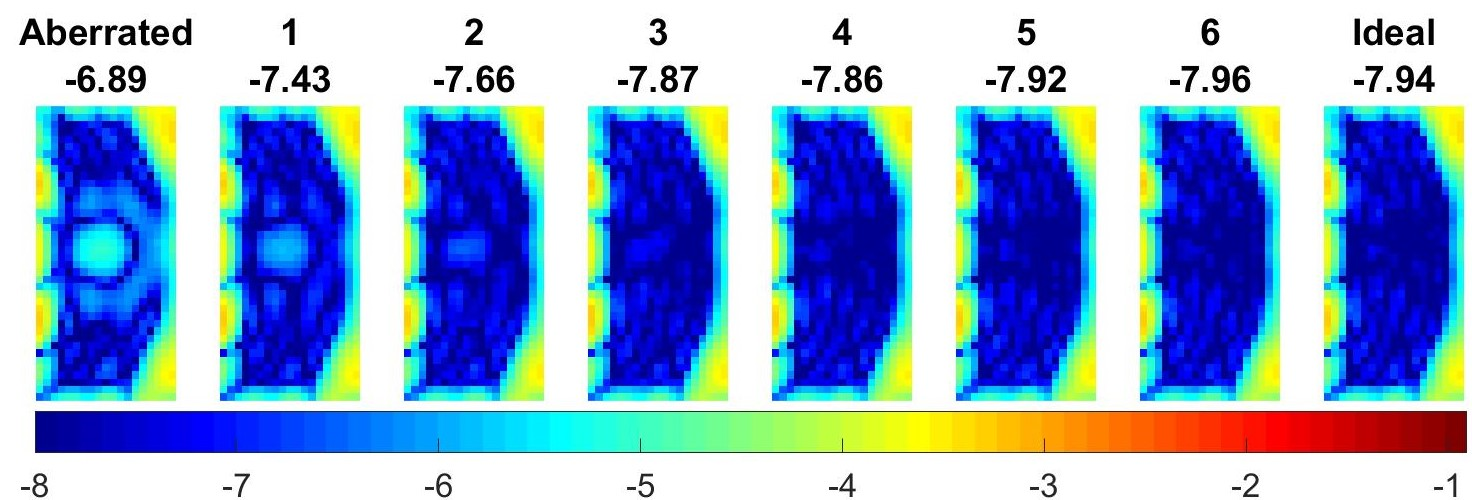}
\caption{Evolution of the DF over the 6 iterations (seen in Fig \ref{fig:speckle_contrast}b) to converge from a degraded DF average contrast of $10^{-6.90}$ with a $10^{-5.0}$ magnitude speckle to the LDFC-corrected DF with $10^{-7.94}$ average contrast.  The ideal DF is shown in the final frame for reference.  Scale is log$_{10}$ contrast.}
\label{fig:speckle_DFevolution}
\end{figure}

\begin{figure}[H]
\centering
\qquad
\subfloat[Average contrast across the full DF for the pre-EFC PSF, DF post-EFC (blue), DF post-EFC with injected speckle (red), and the corrected DF post-LDFC (black)]{\includegraphics[width=0.32\textwidth]{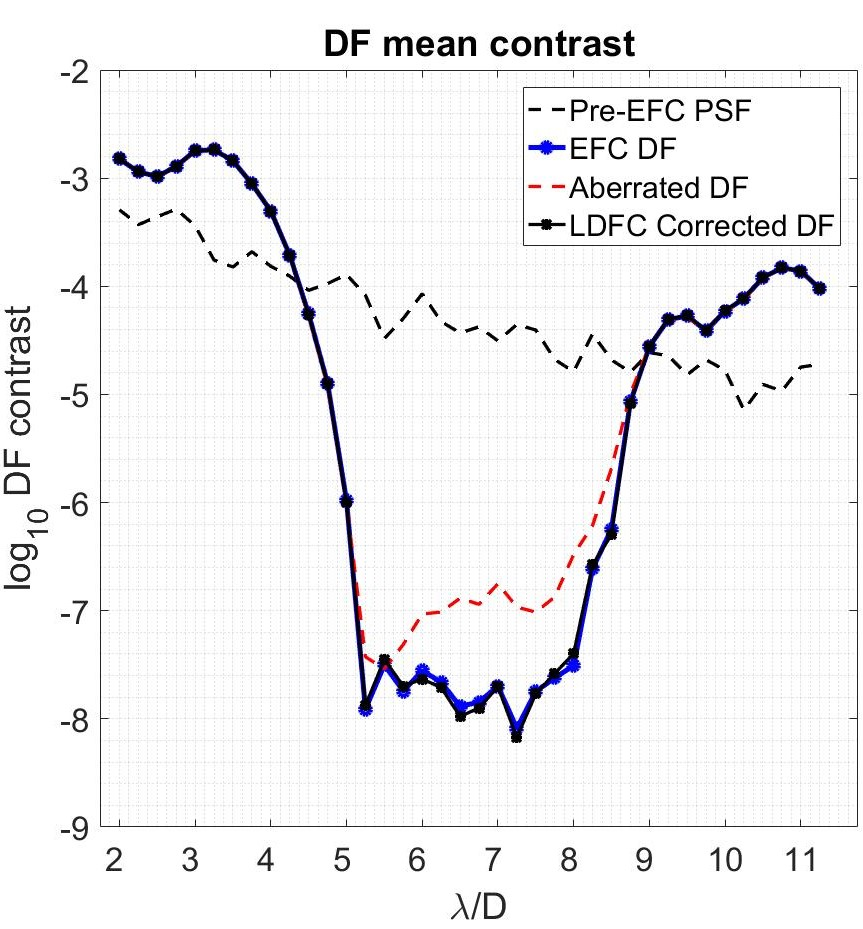}}
\qquad
\subfloat[Average DF contrast (black) over 50 iterations showing convergence to the initial EFC contrast (blue) after 6 iterations.]{\includegraphics[width=0.59\textwidth]{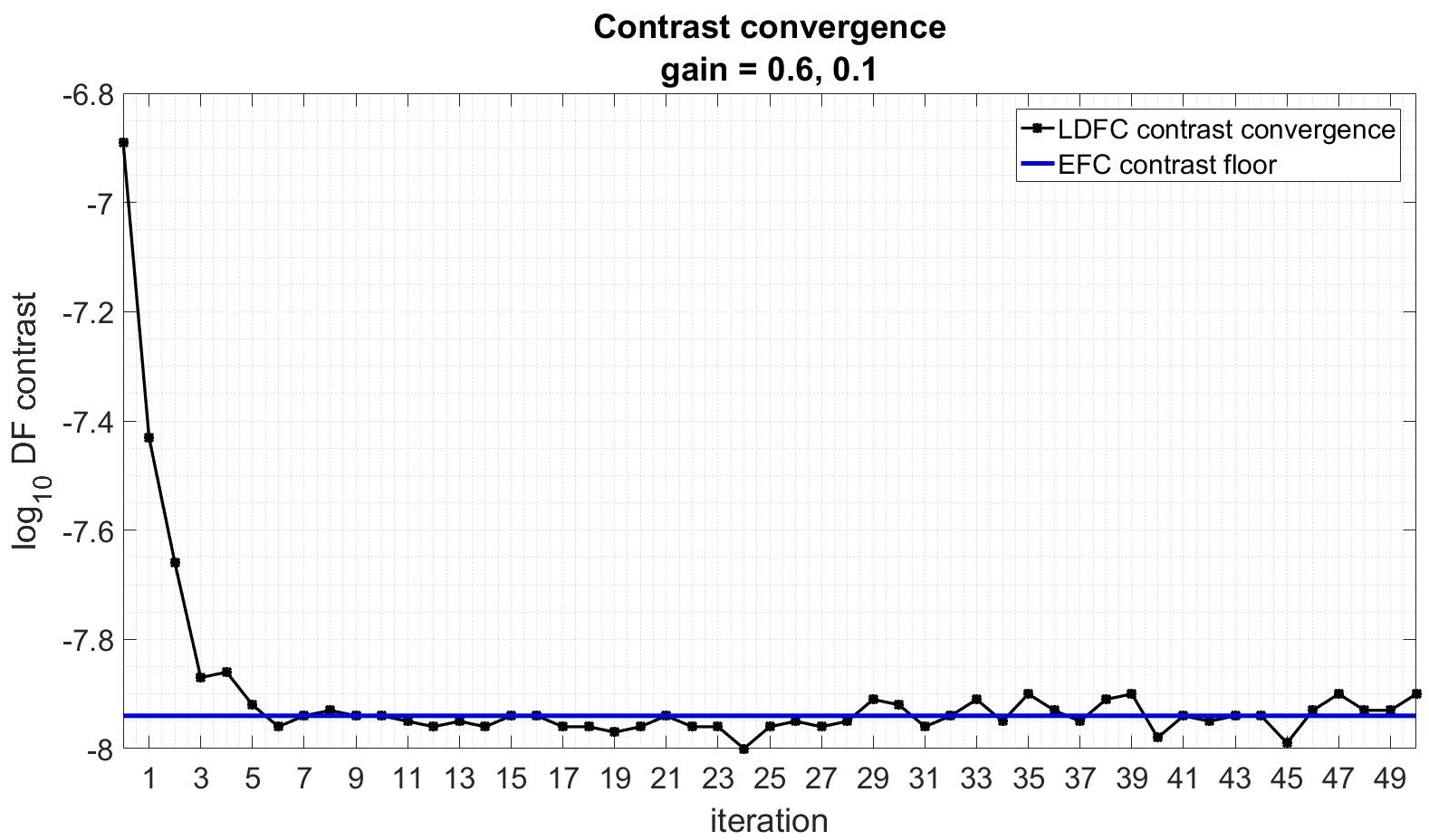}}
\caption{Performance of the spatial LDFC servo with a sinusoidal phase perturbation.  Gain = 0.6 until the DF contrast reached 10$^{-7.9}$.  The gain was lowered to 0.1 for the remaining iterations.}
\label{fig:speckle_contrast}
\end{figure}

After the DF was constructed, a spatial sine wave phase perturbation with 6 cycles/aperture was introduced into the pupil plane, forming a speckle at +/- 6 $\lambda$/D: one speckle within the DF and one speckle within the BF.  The sine perturbation was given a 1 nm peak-to-valley (P-V) amplitude in phase, creating a speckle pair with a maximum magnitude of 10$^{-5.0}$ and an average aberrated DF contrast of 10$^{-6.90}$.  The LDFC control loop was run with a gain of 0.6 until the average DF contrast reached 10$^{-7.9}$ at which point the gain was reduced to 0.1 to maintain the correction.  The LDFC control loop was allowed to run for 50 iterations for this demonstration with convergence occuring after 6 iterations.  The results are shown in Table \ref{tabular:speckletest} and in Fig(s) \ref{fig:speckle_full_display_PUPILs} - \ref{fig:speckle_contrast}.  It should be noted that, in Figs \ref{fig:speckle_DFevolution} and \ref{fig:speckle_contrast}b, the LDFC-corrected DF contrast occasionally drops below the initial EFC contrast level.  This effect is due to noise fluctuations.\\
\begin{table}[H]
    \begin{center}
     \scriptsize
    \renewcommand{\arraystretch}{1.5}
    \begin{tabular}{| >{\centering\arraybackslash}m{1.5in} | >{\centering\arraybackslash}m{1.5in} |}
    \hline
    { EFC DF contrast} & { 10$^{-7.94}$}   \\[0ex] \hline
    { Speckle magnitude} & {10$^{-5.0}$}   \\[0ex] \hline
    { Avg DF contrast with speckle} & {10$^{-6.90}$}   \\[0ex] \hline
    { LDFC DF contrast} & {10$^{-7.94}$}   \\[0ex] \hline
    { $\Delta$Contrast} & {10$^{-1.04}$}   \\[0ex] \hline
    { \# Iterations to converge} & {6}   \\[0ex] \hline
  \end{tabular}
  \caption{Performance with a sine wave phase: Initial EFC DF average contrast, magnitude of the injected speckle, average contrast of the aberrated DF, average contrast of the DF after LDFC, total change in contrast for one full LDFC loop, and the number of iterations to converge to the EFC contrast floor}
  \label{tabular:speckletest}
  \end{center}
\end{table}
%------------------------------------------------
\subsection{Kolmogorov phase perturbation} \label{sec:kol}
\begin{figure}[H]
\centering
\includegraphics[width = 0.75\textwidth]{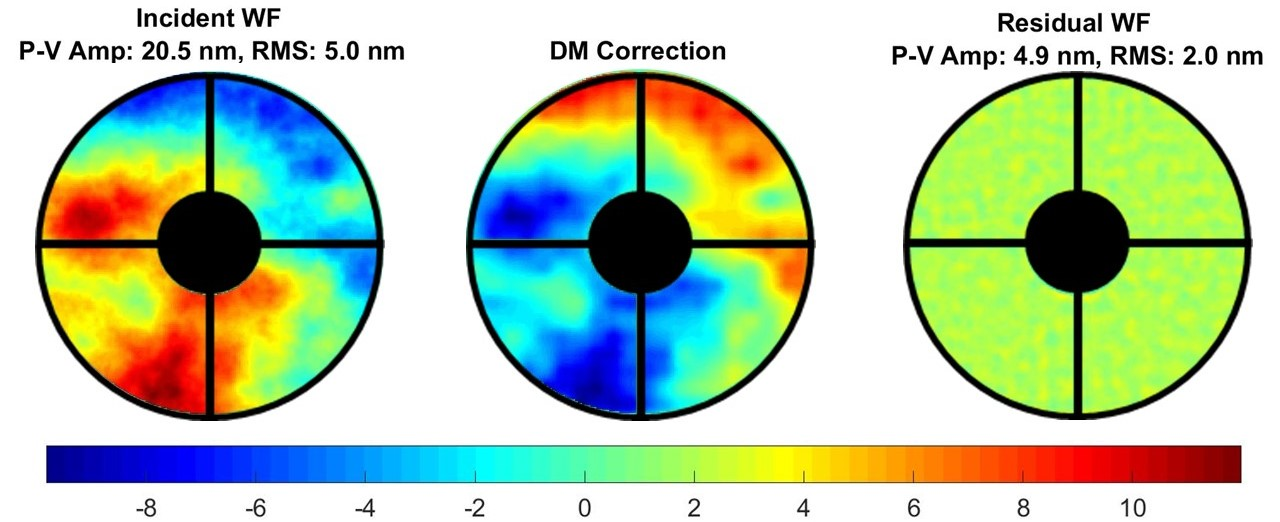}
\caption{The injected Kolmogorov phase perturbation with a P-V amplitude of 20.5 nm (left), the DM response derived by LDFC (center) and the final residual wavefront error (right) after 6 iterations.  Scale is given in nm.}
\label{fig:kol_full_display_PUPILs}
\end{figure}

\begin{figure}[H]
\centering
\includegraphics[width = 0.75\textwidth]{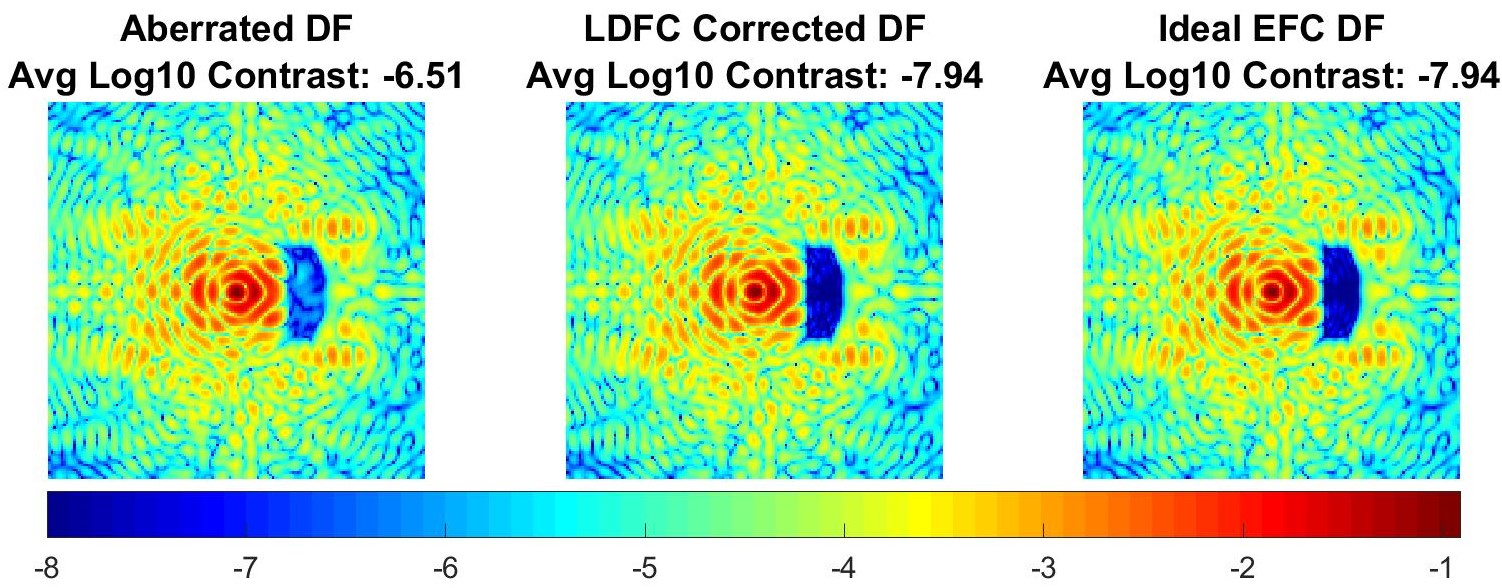}
\label{fig:kol_full_display_PSFs}
\caption{The aberrated PSF with multiple speckles in the DF and average DF contrast of 10$^{-6.51}$, the final LDFC-corrected DF with 10$^{-7.94}$ average DF contrast, and the reference EFC-derived DF with 10$^{-7.94}$ average DF contrast.  Scale is log$_{10}$ contrast.}
\end{figure}

\begin{figure}[H]
\centering
\includegraphics[width=0.75\textwidth]{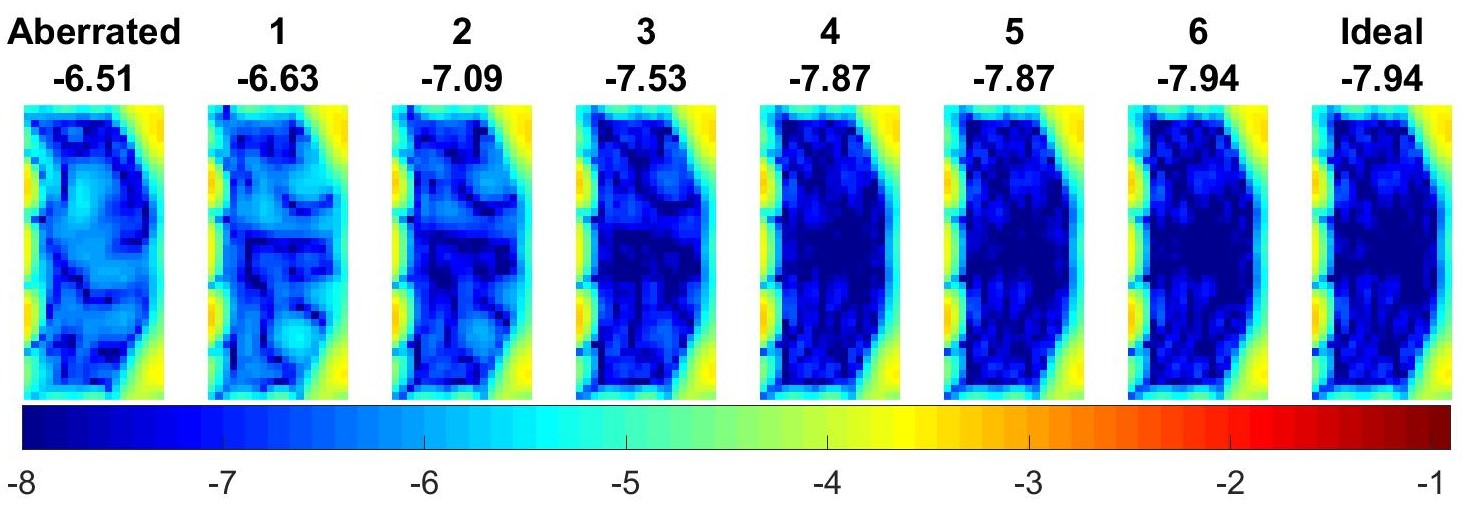}
\caption{Evolution of the DF over the 6 iterations (seen in Fig \ref{fig:kol_contrast}b) to converge from a degraded DF average contrast of $10^{-6.51}$ to the LDFC-corrected DF with $10^{-7.94}$ average contrast.  The ideal DF is shown in the final frame for reference.  Scale is log$_{10}$ contrast.}
\label{fig:kol_DFevolution}
\end{figure}

\begin{figure}[H]
\centering
\qquad
\subfloat[Average contrast across the full DF for the pre-EFC PSF, DF post-EFC (blue), DF post-EFC with injected speckle (red), and the corrected DF post-LDFC (black)]{\includegraphics[width=0.32\textwidth]{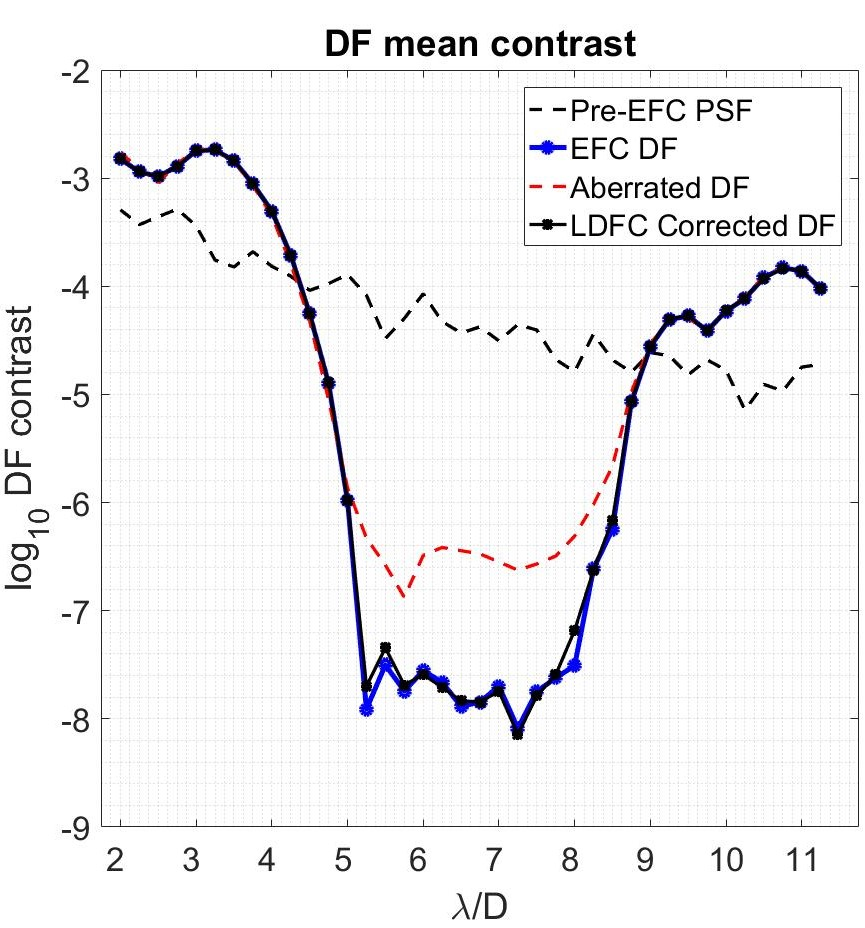}}
\qquad
\subfloat[Average DF contrast (black) over 50 LDFC iterations showing convergence to the initial EFC contrast (blue) after 6 iterations.]{\includegraphics[width=0.59\textwidth]{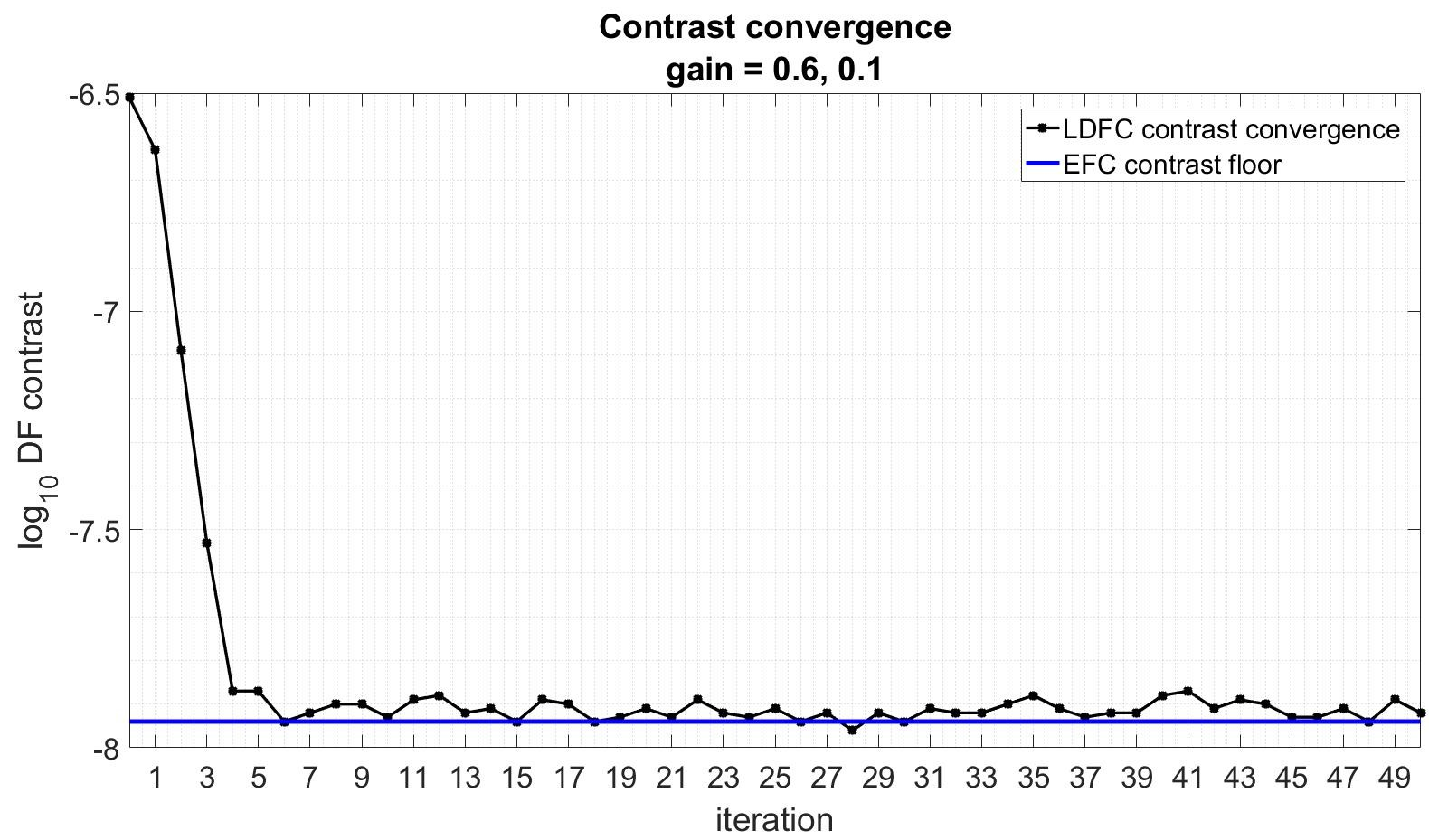}}
\caption{Performance of the spatial LDFC servo with a Kolmogorov phase perturbation.  Gain = 0.6 until the DF contrast reached 10$^{-7.9}$.  The gain was lowered to 0.1 for the remaining iterations.}
\label{fig:kol_contrast} 
\end{figure}

In the first case, the injected aberration created a single speckle in the DF and a corresponding speckle in the BF.  To demonstrate spatial LDFC's ability to suppress multiple speckles, a Kolmogorov phase aberration was generated in the pupil plane instead of a sinusoidal phase perturbation (see Fig \ref{fig:kol_full_display_PUPILs}).   The phase perturbation was given a P-V amplitude of 20.5 nm, creating an aberrated DF with an average contrast of 10$^{-6.51}$.  The LDFC control loop was again run with a gain of 0.6 until the average DF contrast reached 10$^{-7.9}$ at which point the gain was reduced to 0.1 to maintain the correction.  The LDFC control loop was allowed to run for 50 iterations for this demonstration with convergence occuring after 6 iterations.  The results are shown in Table \ref{tabular:koltest} and shown in Fig(s) \ref{fig:kol_full_display_PUPILs} - \ref{fig:kol_contrast}.  As in the previous single speckle demonstration, the LDFC-corrected DF contrast occasionally drops below the initial EFC contrast level in Fig \ref{fig:kol_contrast}b.  This effect is due to noise fluctuations. \\ 
\begin{table}[H]
    \begin{center}
     \scriptsize
    \renewcommand{\arraystretch}{1.5}
    \begin{tabular}{| >{\centering\arraybackslash}m{1.8in} | >{\centering\arraybackslash}m{1.5in} |}
    \hline
    { EFC DF contrast} & { 10$^{-7.94}$}   \\[0ex] \hline
    { Avg DF contrast with aberration} & {10$^{-6.51}$}   \\[0ex] \hline
    { LDFC DF contrast} & {10$^{-7.94}$}   \\[0ex] \hline
    { $\Delta$Contrast} & {10$^{-1.43}$}   \\[0ex] \hline
    { \# Iterations to converge} & {6}   \\[0ex] \hline
  \end{tabular}
  \caption{Performance with Kolmogorov phase: Initial EFC DF average contrast, magnitude of the injected speckle, average contrast of the aberrated DF, average contrast of the DF after LDFC, total change in contrast for one full LDFC loop, and the number of iterations to converge to the EFC contrast floor}
  \label{tabular:koltest}
  \end{center}
\end{table}
%------------------------------------------------
\section{Discussion} \label{sec:discuss}
We have demonstrated here that spatial LDFC is capable of locking the DF contrast at its ideal EFC state using only the BF response to a perturbation in the optical path.  However, there are limitations to spatial LDFC and a potential null space which need to be explored.  These issues and some potential solutions are addressed below.   \\

\subsection{Limitations of spatial LDFC}\label{sec:limits}
One significant limiting factor for spatial LDFC is DF symmetry.  This technique requires access to a BF that is located spatially opposite the DF.  Due to this requirement, spatial LDFC is expected to work only with a non-symmetric DF.  However, in the case of a symmetric DF, spectral LDFC offers a possible solution (see Section \ref{sec:spectralLDFC}).   Since spectral LDFC relies on speckles that are located spatially within the DF but outside of the control bandwidth, it is not affected by the lack of a BF spatially opposite the DF.  In the case of a much larger DF than the one presented here, spatial LDFC is predicted to still be capable of stabilizing the DF, but it cannot use BF speckles at spatial frequencies higher than the those present in the DF to do so.  \\  

In future work, spatial LDFC's null space must also be further explored.  For this technique, the null space will consist of wavefront errors that affect the DF without changing the BF.  One potential example of this null space is the formation of a speckle on a single side of the focal plane due to the combination of phase and amplitude sine wave aberrations.  In such a case, if the speckle falls inside the DF, the BF will not see any modulation and will therefore be unable to sense and correct the aberration.  A second potential null space example would consist of an incident phase aberration sine wave with a phase that creates a BF speckle with a phase that is 90$^{o}$ from the local BF phase.  This case would not create a linear signal and would therefore not be corrected by LDFC.  To date, the simulations presented here have not revealed a null space, but it can exist under system-specific conditions.   It should also be noted that this paper has specifically explored a system in which aberrations were introduced and corrected by the same DM; in real systems there will be aberrations that occur outside the DM-conjugate pupil plane and subsequently do not correspond exactly to DM authority.  Such cases will require further investigation as spatial LDFC continues to develop.     

\subsection{Addressing the null space with spectral LDFC}\label{sec:spectralLDFC}
A potential solution for overcoming spatial LDFC's null space is to operate a separate version of LDFC simultaneously.  This second version, known as spectral LDFC, freezes the state of the DF within the control bandwidth by using state measurements of light outside of the control bandwidth.  This method exploits the fixed wavelength relationships that exist between speckles at different wavelengths that were generated by the same aberration.  To first order, this fixed relationship scales the speckle separation linearly with wavelength and scales the complex amplitude inversely with wavelength.  The complex amplitude speckle field may also interfere with static chromatic coronagraph residuals  due to the coronagraph's finite design bandwidth.  These relationships between out-of-band and in-band light allow for the state of the DF within the control band to be monitored and maintained by measurements  made of speckles located outside of the spectral control band.\cite{Guyon2017_spectralLDFC}  Since spectral and spatial LDFC rely on a BF signal from separate dimensions, the null spaces of the two forms of LDFC are not expected to overlap.  For this reason, concurrent operation of spectral and spatial LDFC can provide a powerful tool for compensating for the separate null spaces of both techniques.\\

\begin{figure}[H]
\centering
\includegraphics[width=0.71\textwidth]{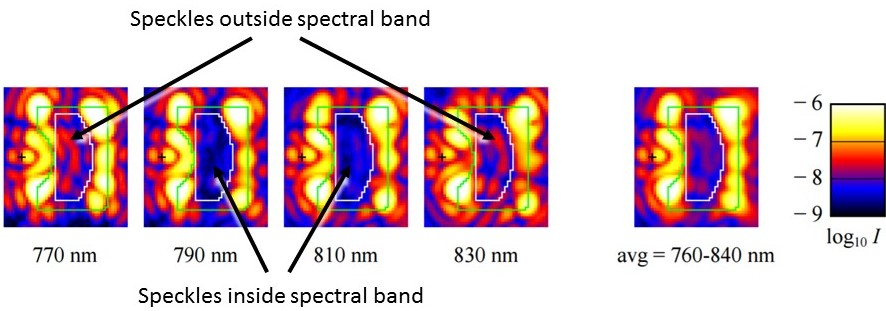}
\caption{Spectral LDFC example: The 7 $\lambda$/D x 8 $\lambda$/D DF from JPL's High Contrast Imaging Testbed (HCIT) created using a PIAA coronagraph.  The DF is shown in four individual spectral channels: two channels within the control bandwidth and two out-of-band channels with speckles used to maintain the DF state within the control bandwidth.  Also shown is the average DF over the full 10\% bandwidth centered at $\lambda$ = 800 nm.\cite{Guyon2009_PIAA_DH}}
\label{fig:spectralLDFC}
\end{figure}

As an example of the BF signal that can be used by spectral LDFC, Fig \ref{fig:spectralLDFC} shows the DF created using a Phase Induced Amplitude Apodization (PIAA)\cite{Guyon2006_PIAA} coronagraph at JPL's High Contrast Imaging Testbed (HCIT).   Speckles within the DF are shown at multiple wavelengths, both in-band (the science image) and out-of-band (the signal used by spectral LDFC). \\
While spectral LDFC was not in operation when this image was taken, this is a clear demonstration of a case in which the in-band DF contrast could be maintained by sensing the speckles that are outside the control bandwidth and applying the appropriate wavelength-scaled correction to cancel the in-band speckles.  Further development and analysis of this form of LDFC can be found in an upcoming paper by Guyon et.al.\cite{Guyon2017_spectralLDFC} \\
\section{Conclusion}\label{sec:conclusion}
In summary, spatial LDFC acts as an extension of EFC by operating as a servo that can maintain high contrast in the DF during science exposures.  Using changes in the BF to provide updates on the state of the field within the DF, spatial LDFC is able to lock the state of the DF after it is established by EFC without relying on field modulation which interrupts the science acquisition and fundamentally limits the exposure time.  The substantial increase in uninterrupted observation time spatial LDFC offers makes it a more efficient method than EFC for maintaining deep contrast and will lead to an overall increase in the number of planets detected and analyzed over the lifetime of an instrument.  Here we have introduced the mathematical principles behind spatial LDFC and provided demonstrations of its capabilities through numerical simulation; future work will include further analysis of this technique's abilities as well as laboratory demonstrations.      

\section{Acknowledgments}
This work was funded by the NASA Exoplanet Exploration Program Segmented Coronagraph Design and Analysis study effort.  The authors would like to acknowledge Dr. Sandrine Thomas for her assistance in implementing EFC for this work.  For their continued support, the authors also acknowledge fellow members of the OG and JRM research group: Justin Knight, Jennifer Lumbres, Alexander Rodack, and Lauren Schatz. \\   
%\pagebreak
%----------------------------------------------------------------------------------------
%	REFERENCE LIST
%----------------------------------------------------------------------------------------
\bibliography{DissertationBib}
\bibliographystyle{spiebib}

%----------------------------------------------------------------------------------------

\end{document}